\documentclass[12pt]{article}
\usepackage{a4,graphicx,amssymb}
\usepackage[small,bf]{caption}

\newcommand{\note}[1]{}                    

\newcommand{\third}{\mbox{\small $\frac{1}{3}$}}         
\newcommand{\twothird}{\mbox{\small $\frac{2}{3}$}}      
\newcommand{\R}{\mbox{\tiny $R$}}                        
\newcommand{\Si}{\mbox{\tiny $S$}}                       
\newcommand{\NS}{\mbox{\tiny $N\!S$}}                    

\def\lsim{\mathrel{\rlap{\lower4pt\hbox{\hskip1pt$\sim$}}
    \raise1pt\hbox{$<$}}}                
\def\gsim{\mathrel{\rlap{\lower4pt\hbox{\hskip1pt$\sim$}}
    \raise1pt\hbox{$>$}}}                

\begin{document}

\title{
\vspace{-3.0cm}
\flushright{\normalsize DESY 11-168} \\
\vspace{-0.35cm}
{\normalsize Edinburgh 2011/27} \\
\vspace{-0.35cm}
{\normalsize Liverpool LTH 922} \\
\vspace{-0.35cm}
{\normalsize February 24, 2012} \\
\vspace{0.5cm}
\centering{\Large \bf Hyperon sigma terms for $2+1$ quark flavours}}

\author{\large
        R. Horsley$^a$, Y. Nakamura$^b$, H. Perlt$^c$, \\
        D. Pleiter$^{de}$, P.~E.~L. Rakow$^f$, G. Schierholz$^{eg}$, \\
        A. Schiller$^c$, H. St\"uben$^h$, F. Winter$^a$ \\
        and J.~M. Zanotti$^a$ \\[1em]
         -- QCDSF-UKQCD Collaboration -- \\[1em]
        \small $^a$ School of Physics and Astronomy,
               University of Edinburgh, \\[-0.5em]
        \small Edinburgh EH9 3JZ, UK \\[0.25em]
        \small $^b$ RIKEN Advanced Institute for
               Computational Science, \\[-0.5em]
        \small Kobe, Hyogo 650-0047, Japan \\[0.25em]
        \small $^c$ Institut f\"ur Theoretische Physik,
               Universit\"at Leipzig, \\[-0.5em]
        \small 04109 Leipzig, Germany \\[0.25em]
        \small $^d$ JSC, J\"ulich Research Centre, \\[-0.5em]
        \small 52425 J\"ulich, Germany \\[0.25em]
        \small $^e$ Institut f\"ur Theoretische Physik,
               Universit\"at Regensburg, \\[-0.5em]
        \small 93040 Regensburg, Germany \\[0.25em]
        \small $^f$ Theoretical Physics Division,
               Department of Mathematical Sciences, \\[-0.5em]
        \small University of Liverpool,
               Liverpool L69 3BX, UK \\[0.25em]
        \small $^g$ Deutsches Elektronen-Synchrotron DESY, \\[-0.5em]
        \small 22603 Hamburg, Germany \\[0.25em]
        \small $^h$ Konrad-Zuse-Zentrum
               f\"ur Informationstechnik Berlin, \\[-0.5em]
        \small 14195 Berlin, Germany }

\date{}

\maketitle



\begin{abstract}
   QCD lattice simulations determine hadron masses
   as functions of the quark masses. From the gradients of
   these masses and using the Feynman--Hellmann theorem 
   the hadron sigma terms can then be determined. We use here
   a novel approach of keeping the singlet quark mass constant in our
   simulations which upon using an $SU(3)$ flavour symmetry breaking
   expansion gives highly constrained (i.e.\ few parameter)
   fits for hadron masses in a multiplet. This is a highly
   advantageous procedure for determining the hadron mass gradient
   as it avoids the use of delicate chiral perturbation theory.
   We illustrate the procedure here by estimating the
   light and strange sigma terms for the baryon octet.
\end{abstract}




\section{Introduction} 


Hadron sigma terms, $\sigma_l^{(H)}$, $\sigma_s^{(H)}$ are defined%
\footnote{Or more accurately as the matrix element of the double
commutator of the Hamiltonian with two axial charges.
However this is equivalent to the definition given
in eq.~(\ref{sig_def}), see for example \cite{cheng88a}.}
as that part of the mass of the hadron (for example the nucleon)
coming from the vacuum connected expectation value of the up ($u$)
down ($d$) and strange ($s$) quark mass terms in the QCD Hamiltonian,
\begin{eqnarray}
   \sigma_l^{(H)} = m_l^{\R}\langle H|(\overline{u}u + \overline{d}d)^{\R} 
                                   |H\rangle \,,
   \qquad
   \sigma_s^{(H)} = m_s^{\R}\langle H|(\overline{s}s)^{\R} 
                                   |H\rangle \,,
\label{sig_def}
\end{eqnarray}
where we have taken the $u$ and $d$ quarks to be mass degenerate,
$m_u = m_d \equiv m_l$. (The superscript $^{\R}$ denotes a renormalised
quantity.) Other contributions to the hadron mass
come from the chromo-electric and chromo-magnetic gluon pieces
and the kinetic energies of the quarks, \cite{ji94a}.
Sigma terms are interesting because they are sensitive
to chiral symmetry breaking effects. Experimentally the value
for $\sigma_l^{(N)}$ has been deduced from low energy
$\pi$-$N$ scattering. A delicate extrapolation to the
chiral limit \cite{cheng88a} gives a result for the isospin even
amplitude of $\sigma_{\pi N}/f_\pi^2$ (with
$\sigma_{\pi N} \equiv \sigma_l^{(N)}$), from which the sigma
term may be found. The precise value obtained this way has been
under discussion for many years. However within the limits of our
lattice calculation, this will not concern us here and for
orientation we shall just quote a range of results from
earlier analyses of \cite{koch82a,gasser91a}
of $45(8)\,\mbox{MeV}$ while a later dispersion analysis
\cite{pavan01a} suggested a much higher value $64(7)\,\mbox{MeV}$.
An estimation using heavy baryon chiral perturbation theory
gave $45\,\mbox{MeV}$, \cite{borasoy96a}. 
A more recent estimate gave $59(17)\,\mbox{MeV}$, \cite{martin-camalich10a}. 
Even less is known about the nucleon strange sigma term.
Eq.~(\ref{sig_def}) is usually written (in particular for the nucleon) as
\begin{eqnarray}
   \sigma_l^{(N)} 
      = { m_l^{\R}\langle N|(\overline{u}u + \overline{d}d
                         - 2 \overline{s}s)^{\R} |N \rangle 
        \over 1 - y^{(N)\R} } \,, \qquad
   y^{(N)\R} 
      = { 2 \langle N| (\overline{s}s)^{\R} |N \rangle \over
          \langle N| (\overline{u}u + \overline{d}d)^{\R} |N \rangle } \,,
\label{sig_y_def}
\end{eqnarray}
(i.e.\ we consider $\sigma_l^{(N)}$ and $y^{(N)\R}$ rather than
$\sigma_l^{(N)}$ and $\sigma_s^{(N)}$). The simplest calculation,
e.g.\ \cite{cheng88a} (which we will discuss in more detail later)
uses first order in $SU(3)$ flavour symmetry (octet) breaking to give
\begin{eqnarray}
   \sigma_l^{(N)}
      = { m_l^{\R} \over m_s^{\R} - m_l^{\R} }
           { M_\Xi + M_\Sigma - 2M_N \over 1 - y^{(N)\R} }
              \sim { 26 \over 1 - y^{(N)\R} } \,\mbox{MeV} \,,
\label{sigl_est}
\end{eqnarray}
and
\begin{eqnarray}
   \sigma_s^{(N)}
      = { m_s^{\R} \over m_l^{\R} } {1\over 2}y^{(N)\R} \sigma_l^{(N)}
              \sim 325 {  y^{(N)\R} \over 1 - y^{(N)\R} } \,\mbox{MeV} \,,
\label{sigs_est}
\end{eqnarray}
where $m_s^{\R}/ m_l^{\R}$ is the ratio of the strange to light quark masses,
which using the leading order PCAC formula for this ratio gives
\begin{eqnarray}
   m_s^{\R}/m_l^{\R} = (2M_K^2 - M_\pi^2)/M_\pi^2 \sim 25 \,.
\label{qmrat_est}
\end{eqnarray}
The Zweig rule, $\langle N| (\overline{s}s)^{\R} |N \rangle \sim 0$
would then give
\begin{eqnarray}
   \sigma_l^{(N)} \sim 26\,\mbox{MeV}\,, \qquad
   \sigma_s^{(N)} \sim 0\,\mbox{MeV} \,,
\end{eqnarray}
while any non-zero strangeness content, $y^{(N)\R} > 0$ would increase
this value of $\sigma_l^{(N)}$, $\sigma_s^{(N)}$ (and indeed, due to the
large coefficient, $\sigma_s^{(N)}$ quite rapidly).

Determination of the strange sigma term (and in particular $y^{(N)\R}$)
is important in constraining the cross section for the detection
of dark matter. WIMPs would be scattered off nuclei by the exchange
of scalar particles, such as the Standard Model Higgs particle,
which will interact more strongly with heavier quark flavours.
This coupling can be parameterised in terms of the fractional
contribution of a quark flavour $q$ to the nucleon's mass $M_N$,
$f_{T_q} = m_q^{\R} \langle N|(\overline{q}q)^{\R}|N \rangle / M_N$.
While the contributions of the charm and heavier flavours approach
a constant that is proportional to the gluonic contribution $f_{T_g}$,
there is a strong dependence of the cross section on the value of
$f_{T_s}$ , see e.g.\ \cite{ellis09a,giedt09a} and references therein.

Computing the sigma terms from lattice QCD has a long history
from initial quenched simulations to $2$ flavour and more recently $2+1$
flavour simulations, e.g.\ \cite{gusken88a,altmeyer93a,fukugita94a,dong96a,
gusken98a,walker-loud08a,ohki08a,young09a,ishikawa09a,touissant09a,
takeda10a,durr11a}, with a status report being given in \cite{young09b}.
In general more recent results tend to give a lower $\sigma_s^{(N)}$ term
than earlier determinations.

In this article, we shall investigate this simple picture
as described in eqs.~(\ref{sigl_est}), (\ref{qmrat_est})
and in particular test the linearity assumption of $SU(3)$ flavour
symmetry breaking.


\section{Flavour symmetry expansions}
\label{flavour_sym_expan}


Lattice simulations start at some point in the $(m_s^{\R}, m_l^{\R})$
plane and then approach the physical point $(m_s^{\R\,*}, m_l^{\R\,*})$
along some path. (In future we shall denote the physical point
with a $^*$.) As we shall be considering flavour symmetry breaking
then we shall start here at a point on the flavour symmetric line
$m_l^{\R} = m_s^{\R}$ and then consider the path keeping the average
quark mass constant, $\overline{m} = \mbox{const.}$. The $SU(3)$
flavour group (and quark permutation symmetry) then restricts
the quark mass polynomials that are allowed,
\cite{bietenholz11a}, giving for the baryon octet
\begin{eqnarray}
   M_H = M_0(\overline{m}) + c_H\delta m_l + O(\delta m_l^2) \,,
\label{baryon_octet_linfit}
\end{eqnarray}
with
\begin{eqnarray}
   c_H = \left\{ \begin{array}{cc}
                    3A_1       & H = N       \\
                    3A_2       & H = \Lambda \\
                   -3A_2       & H = \Sigma  \\
                   -3(A_1-A_2) & H = \Xi     \\
                 \end{array}
         \right.
\label{cHfoHinO}
\end{eqnarray}
where
\begin{eqnarray}
   \delta m_l   = m_l - \overline{m}\,, \qquad 
   \overline{m} = \third (2m_l + m_s) \,,
\end{eqnarray}
and $A_1$ and $A_2$ are unknown coefficients. So to linear order
in the quark mass, we only have two unknowns (rather than four).
A similar situation also holds for the pseudoscalar and vector octets
(one unknown) and baryon decuplet (also one unknown). These 
functions highly constrain the numerical fits. (At $O(\delta m_l^2)$
only the baryon decuplet has a further constraint.)

Permutation invariant functions of the masses $X_S$, (or `centre of mass'
of the multiplet) can be defined which have no linear dependence on the
quark mass. For example for the baryon octet we have
\begin{eqnarray}
   X_N = \third( M_N + M_\Sigma + M_\Xi)
           = M_0(\overline{m}) +  O(\delta m_l^2) \,.
\label{XN_def}
\end{eqnarray}
(The corresponding result for the pseudoscalar octet is given
later in eq.~(\ref{pseudoscalar_octet_linfit}).)

Furthermore expanding about a specific fixed point, $m_l = m_s = m_0$
on the flavour symmetric line and allowing $\overline{m}$ to vary,
we then have
\begin{eqnarray}
   M_0(\overline{m}) = M_0(m_0) + M_0^\prime(m_0)(\overline{m}-m_0)
                                + O((\overline{m}-m_0)^2) \,.
\label{M0_expan}
\end{eqnarray}
We will see that $A_1$, $A_2$ give all the non-singlet hyperon
sigma terms and $M^\prime(m_0)$ the singlet terms. 

\begin{figure}[htb]
   \begin{minipage}{0.40\textwidth}
   \includegraphics[width=7.00cm]{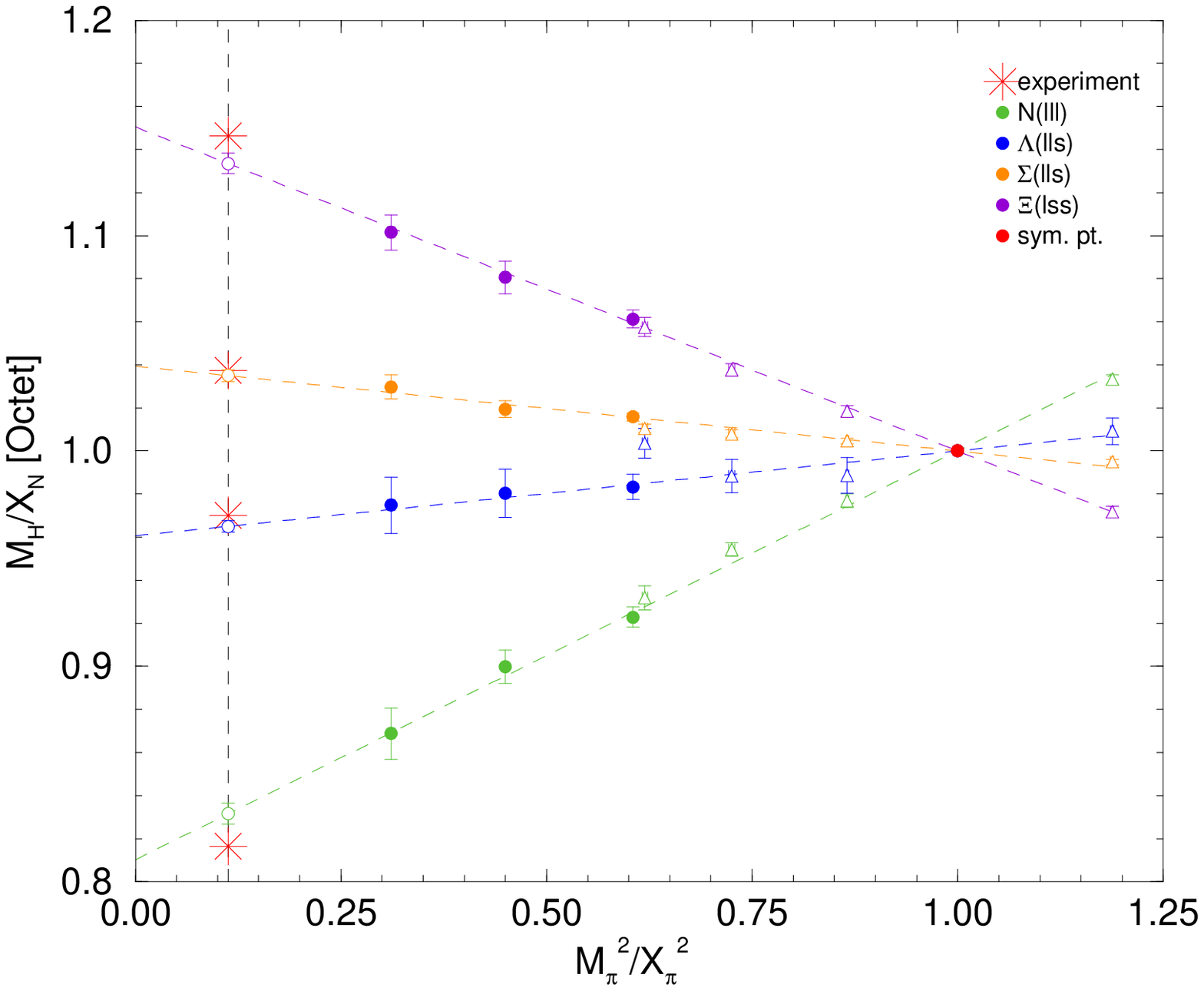}
   \end{minipage} \hspace*{0.10\textwidth}
   \begin{minipage}{0.40\textwidth}
   \includegraphics[width=7.00cm]{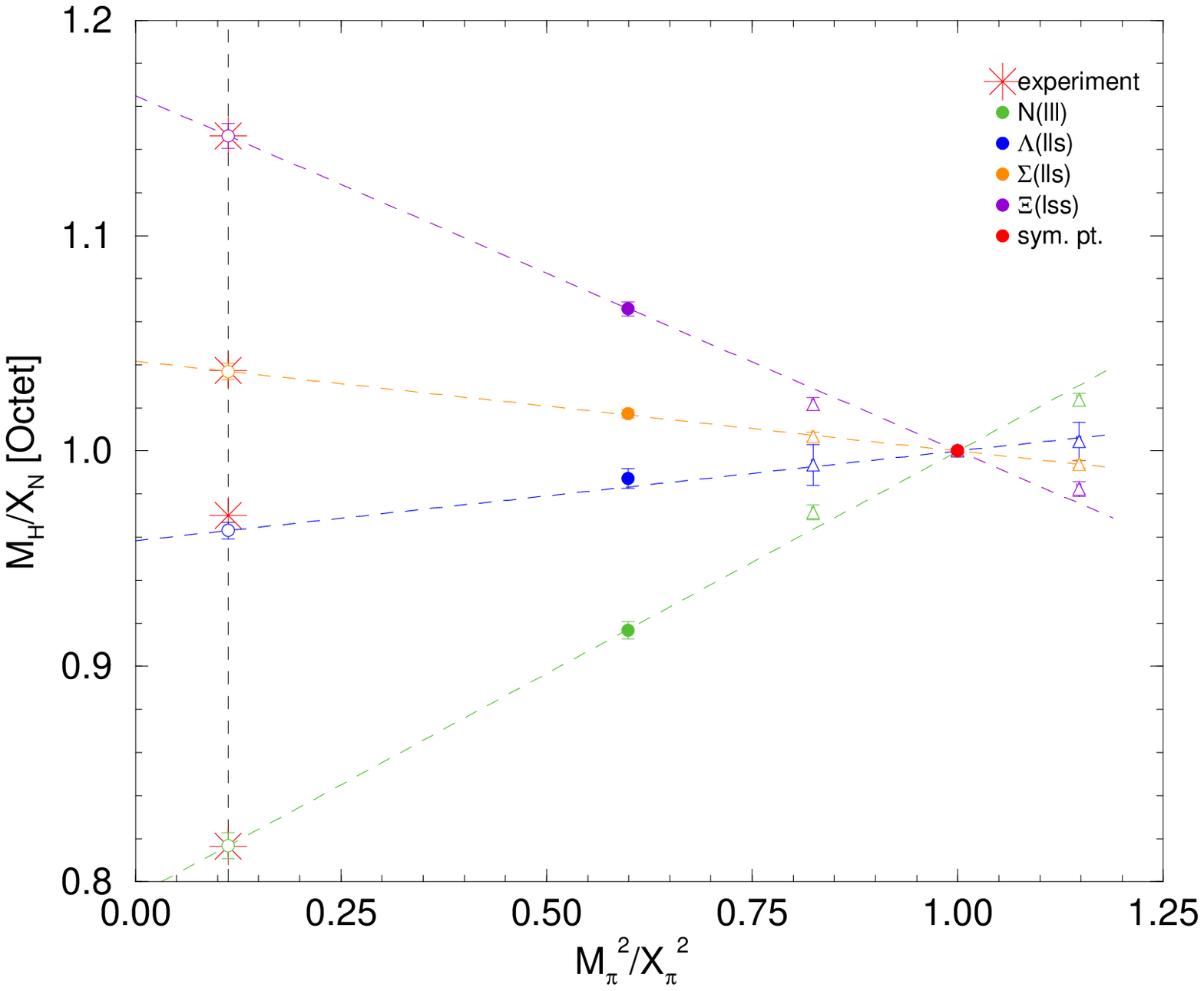}
   \end{minipage}
   \caption{$M_H/ X_N$ ($H = N$, $\Lambda$, $\Sigma$, $\Xi$)
            against $M_\pi^2/X_\pi^2$ for an initial point 
            (``sym. pt.'') on the
            flavour symmetric line given by $\kappa_0 = 0.12090$,
            left panel, and $\kappa_0 = 0.12092$, right panel.
            The $32^3\times 64$ lattices are filled circles,
            while the $24^3\times 48$ lattices are open triangles.
            Also shown is the combined fit of
            eq.~(\protect\ref{mNoXN_mps2oXpi2}) (the dashed lines)
            to the $32^3\times 64$ lattice data. The fit results
            are the open circles, while the experimental points are
            the (red) stars. $l$ and $s$ denote the light and
            strange quark content of the hadron.}
\label{mpsO2omNOpmSigOpmXiOo32_mNOomNOpmSigOpmXiOo3_32x64_lin}
\end{figure}

As an example of the quark mass expansion from a point on the flavour
symmetric line in
Fig.~\ref{mpsO2omNOpmSigOpmXiOo32_mNOomNOpmSigOpmXiOo3_32x64_lin}
we plot the baryon octet $M_H/ X_N$ for $H = N$, $\Lambda$,
$\Sigma$, $\Xi$ against $M_\pi^2/X_\pi^2$ together with a linear
fit, eq.~(\ref{baryon_octet_linfit})
and implicitly eq.~(\ref{pseudoscalar_octet_linfit})
using $2+1$ $O(a)$ improved clover fermions at $\beta = 5.50$,
\cite{cundy09a} using two starting values for the quark mass
on the flavour symmetric line, namely $\kappa_0 = 0.12090$, $0.12092$.

All the points have been arranged in the simulation to have constant
$\overline{m}$. We see that a linear fit provides a good description
of the numerical data from the symmetric point (where
$M_\pi \sim X_\pi^* = 410.9\,\mbox{MeV}$) down to the physical pion mass.

In a little more detail, the bare quark masses are defined as
\begin{eqnarray}
   am_q = {1 \over 2} 
            \left ({1\over \kappa_q} - {1\over \kappa_{0;c}} \right) \,,
            \qquad \mbox{with} \quad q = l, s, 0 \,,
\label{kappa_bare}
\end{eqnarray}
(with the index $q = 0$ denoting the common quark along the flavour
symmetric line) and where vanishing of the quark mass along the $SU(3)$
flavour symmetric line determines $\kappa_{0;c}$. Keeping
$\overline{m} = \mbox{constant} \equiv  m_0$ gives
\begin{eqnarray}
   \kappa_s = { 1 \over { {3 \over \kappa_0} - {2 \over \kappa_l} } } \,.
\label{kappas_mbar_const}
\end{eqnarray}
So once we decide on a $\kappa_l$ this then determines $\kappa_s$.
Note that $\kappa_{0;c}$ drops out of eq.~(\ref{kappas_mbar_const}),
so we do not need its explicit value.
These initial $\kappa_0$ values chosen here, namely $\kappa_0 = 0.12090$
and $0.12092$ are close to the path that leads to the physical point
($\kappa_0 = 0.12092$ being slightly closer). (This is discussed in
more detail in \cite{bietenholz11a}, which also contains
numerical tables and phenomenological values for the hadron masses.
Results not included there are given in Appendix~\ref{hadron_masses}.)
This path is also illustrated later in section~\ref{det_coeff},
Fig.~\ref{mps2oXn2_2mpsK2-mps2oXn2}.
Although finite size effects tend to cancel in ratios of quantities
from the same multiplet, we nevertheless fit just to the results
from the $32^3\times 64$ lattices (filled circles) using the
linear fit of eq.~(\ref{baryon_octet_linfit}). Finally note that
we also have a similar flavour expansion for the pseudoscalar
octet as for the baryon octet, as will be discussed
in section~\ref{det_coeff}.


\section{(Hyperon) scalar matrix elements}
\label{scalar_MEs}

\begin{figure}[htb]
\begin{minipage}{0.40\textwidth}
   \includegraphics[width=7.00cm]{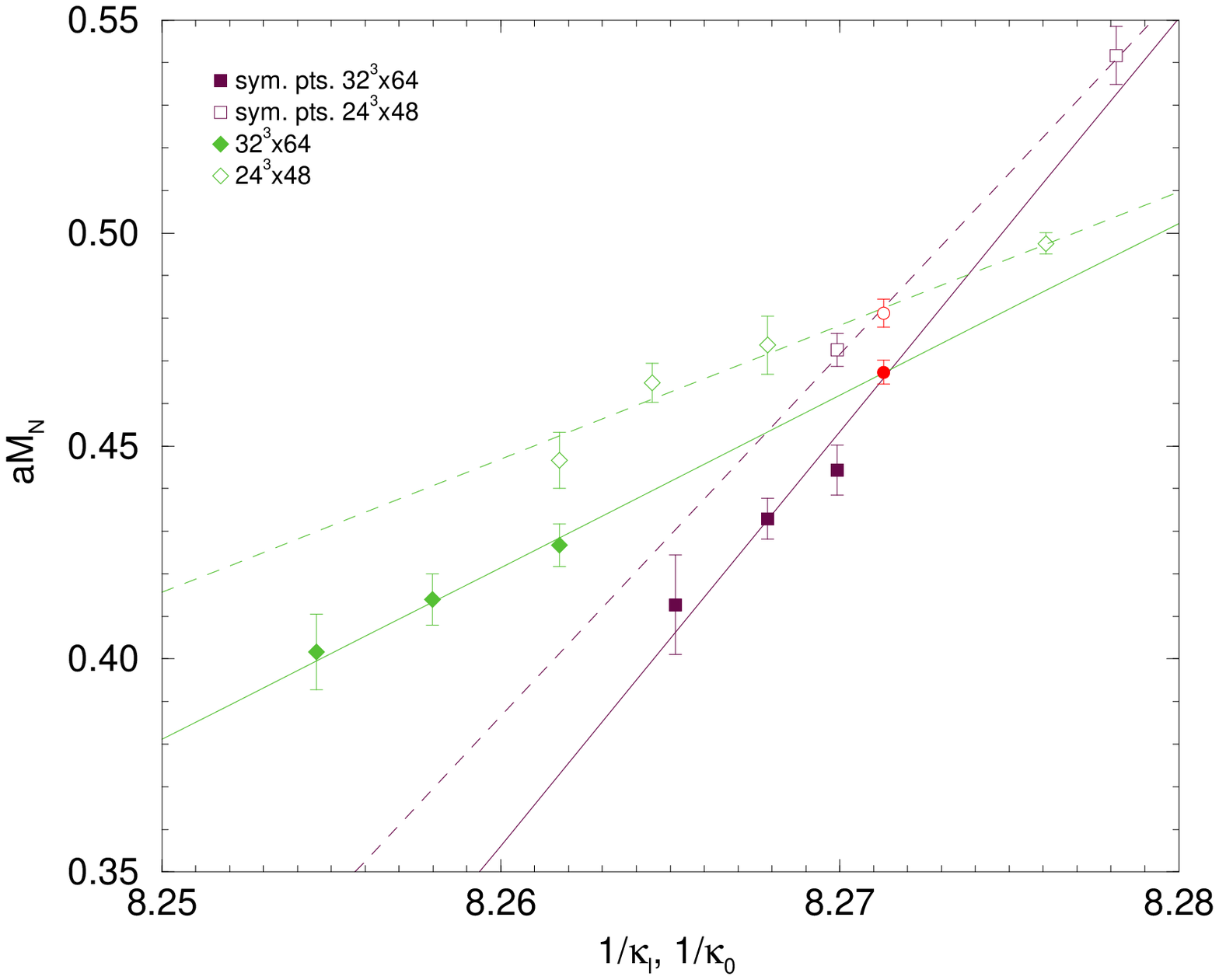}
\end{minipage} \hspace*{0.10\textwidth}
\begin{minipage}{0.40\textwidth}
   \includegraphics[width=7.00cm]{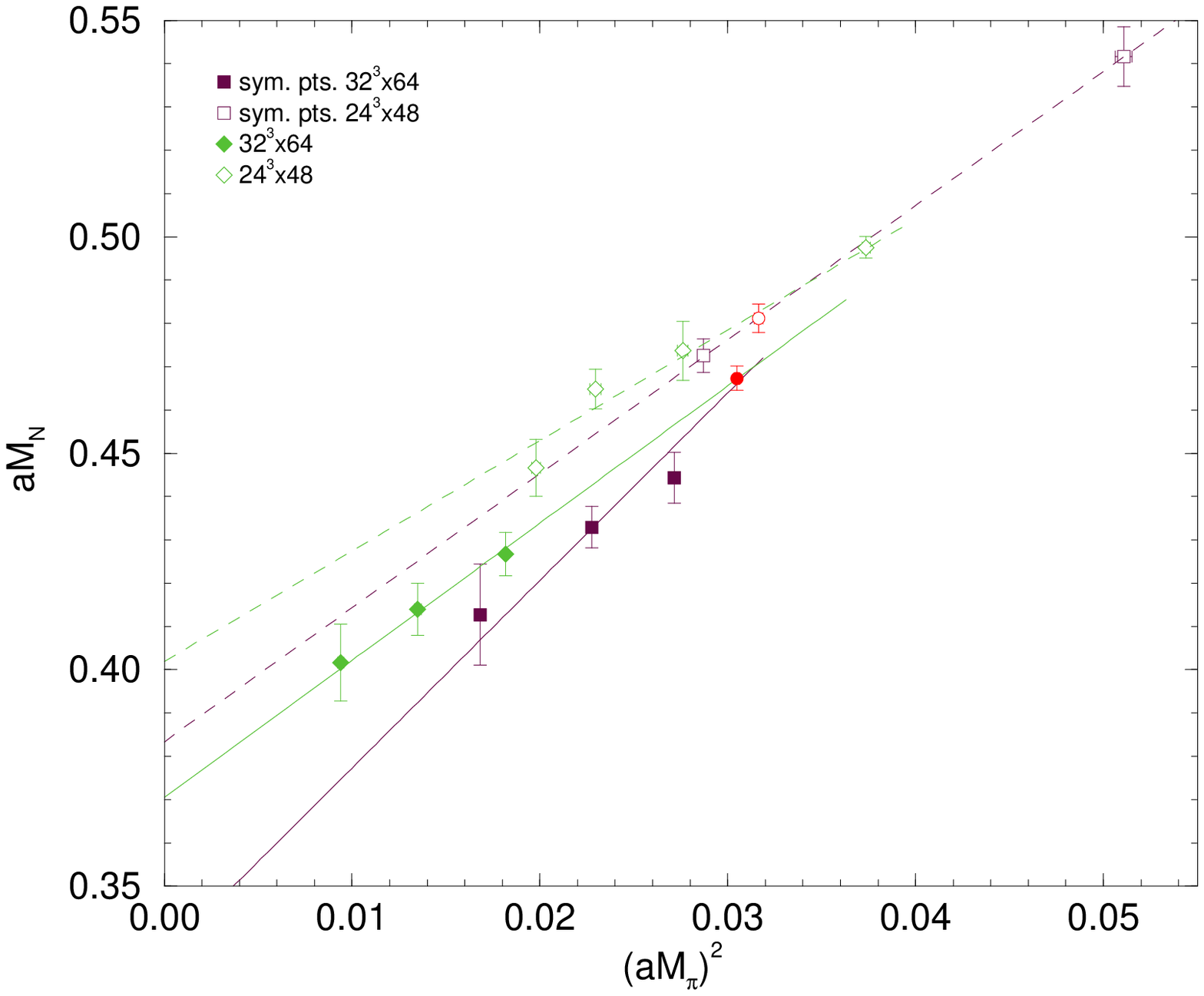}
\end{minipage}
\caption{The left panel shows the nucleon mass, $aM_N$, versus
         $1/\kappa_l$, (for the $\overline{m} = \mbox{const.}$
         points, green diamonds with $\kappa_0 = 0.12090$) and versus
         $1/\kappa_0$ (for the flavour symmetric points, ``sym.\ pts.'',
         maroon squares). The common flavour symmetric points
         are denoted by red circles. The $24^3\times 48$ volume
         results are open symbols together with a dashed line
         for the (linear) fit, while the $32^3\times 64$
         volume results are filled symbols and solid lines.
         Similarly the right panel shows the  nucleon mass $aM_N$,
         versus $(aM_\pi)^2$ (same notation as for the left panel).}
\label{amN_plots}
\end{figure}

Scalar matrix elements can be determined from the gradient of the hadron
mass (with respect to the quark mass) by using the Feynman--Hellman theorem
which is true for both bare and renormalised quantities. So if we take the
derivative with respect to the bare quark mass we get the bare
$\overline{q}q$ matrix element,
\begin{eqnarray}
   {\partial M_H \over \partial m_l } 
      = \langle H| (\overline{u}u + \overline{d}d) | H \rangle\,, \qquad
   {\partial M_H \over \partial m_s } 
      = \langle H| \overline{s}s | H \rangle \,,
\label{fh_thm}
\end{eqnarray}
while if we take the derivative with respect to the renormalised
quark mass we get the renormalised matrix element.
In the left panel of Fig.~\ref{amN_plots}, we show the  
nucleon masses (green diamonds) and the flavour symmetric nucleon
masses (maroon squares) against $1/\kappa_l$, $1/\kappa_0$ respectively
(from eq.~(\ref{kappa_bare}) these are proportional to the
bare quark mass). From the Feynman--Hellmann theorem,
the slope of the masses (maroon squares) gives the total
$\sum_{q=u,d,s} \langle N|\overline{q}q|N\rangle$, while the
slope of the masses (green diamonds) gives the valence contribution%
\footnote{Eq.~(\ref{baryon_octet_linfit}) can be extended
to the `partially quenched' case, \cite{bietenholz11a},
where the sea quark masses remain constrained by
$\overline{m} = \mbox{const.}$ but the valence quark
masses $\mu_l$, $\mu_s$ are unconstrained. Defining
$\delta\mu_q = \mu_q - \overline{m}$ then for the
nucleon, the leading change is particularly simple,
$c_N\delta m_l \to c_N\delta\mu_l$. For the other
members of the octet, $\Lambda$, $\Sigma$, $\Xi$,
both $\delta\mu_l$, $\delta\mu_s$ occur, \cite{bietenholz11a}.}.
The difference between the two contributions gives the disconnected
contribution. Because here all three quark masses are equal, the
disconnected contribution for all three quarks will be the same.
The two slopes thus give the estimates
\begin{eqnarray}
   {\sum_q \langle N|\overline{q}q|N\rangle_{con} \over
      \sum_q \langle N|\overline{q}q|N\rangle} 
      &\sim& {4.0 \over 9.7} \sim 0.41
                                                             \nonumber   \\
   { \langle N|\overline{s}s|N\rangle \over
      \sum_q \langle N|\overline{q}q|N\rangle} 
      &\sim& {1\over 3}\left( 9.7 - 4.0 \over 9.7 \right)
             \sim 0.19 \,,
\end{eqnarray}
for bare lattice quantities.

To look at renormalised matrix elements, we need a plot 
against the renormalised mass, $(aM_\pi)^2$ (as in leading
order PCAC, $M_\pi^2$ is proportional to the renormalised
quark mass, eq.~(\ref{mpi2_pcac})). This is shown in the right
panel of Fig.~\ref{amN_plots}. The slopes are now much closer
to each other. We now find the estimates
\begin{eqnarray}
   {\sum_q \langle N|(\overline{q}q)^{\R}|N\rangle_{con} \over
      \sum_q \langle N|(\overline{q}q)^{\R}|N\rangle} 
      &\sim& {3.2 \over 4.3} \sim 0.74
                                                             \nonumber   \\
   { \langle N|(\overline{s}s)^{\R}|N\rangle \over
      \sum_q \langle N|(\overline{q}q)^{\R}|N\rangle} 
      &\sim& {1\over 3}\left( 4.3 - 3.2 \over 4.3 \right)
      \sim 0.085 \,,
\end{eqnarray}
for renormalised lattice quantities, giving 
$y^{(N)\R} \sim 2 \times 0.085 / (1-0.085) \sim 0.19$. So although for bare
matrix elements, there is a significant strange quark content
this is reduced in the renormalised matrix element.

We shall now try to make these considerations a little more quantitive.


\section{(Hyperon) $\sigma$ equations}
\label{hyperon_sigma_eqns}


\subsection{Renormalisation}


For Wilson (clover) fermions under renormalisation the singlet and
non-singlet pieces of the quark mass renormalise differently
\cite{gockeler04a,rakow04a}. We have
\begin{eqnarray}
   m_q^{\R} = Z^{\NS}\left[m_q + \alpha_Z\third(2m_l+m_s)\right]\,,
             \qquad \alpha_Z = { Z^{\Si} - Z^{\NS} \over Z^{\NS}} \,.
\label{renorm_mq}
\end{eqnarray}
In the action the term
$\sum_q m_q \overline{q}q = \sum_q m_q^{\R} (\overline{q}q)^{\R}$ i.e.\
a renormalisation group invariant or RGI quantity. Upon writing this
in a matrix form and inverting gives
\begin{eqnarray}
   (\overline{q}q)^{\R} 
     = {1 \over Z^{\NS}}
          \left[ \overline{q}q 
                 - {\alpha_Z \over 1+\alpha_Z}
                      \third(\overline{u}u+\overline{d}d+\overline{s}s)
          \right] \,,
\label{qqbar_ren}
\end{eqnarray}
so for $\alpha_Z \not=0$ then there is always mixing between bare operators.

As an example of where this manifests itself, the relation between
the bare, $y^{(H)}$, and renormalised $y^{(H)\R}$, cf.\ eq.~(\ref{sig_y_def}),
is then given by
\begin{eqnarray}
   y^{(H)\R} = { y^{(H)} - \twothird\alpha_Z(1- y^{(H)}) \over
                    1 + \third\alpha_Z(1- y^{(H)}) } \,,
\end{eqnarray}
so we see that $y^{(H)\R} \not= y^{(H)}$ for clover fermions.
Additionally, since $\alpha_Z > 0$ and $y^{(H)} \gsim 0$ we find that 
$y^{(H)\R} < y^{(H)}$, i.e.\ is reduced.

Useful quark combinations are the octet and singlet combinations, namely
\begin{eqnarray}
   (\overline{u}u+\overline{d}d)^{\R} - 2(\overline{s}s)^{\R}
    &=& {1 \over Z^{\NS}}\,
         \left[ (\overline{u}u+\overline{d}d) - 2(\overline{s}s) \right] \,,
                                                             \nonumber   \\
   (\overline{u}u+\overline{d}d)^{\R} + (\overline{s}s)^{\R}
    &=& {1 \over Z^{\NS}(1+\alpha_Z)}\,
         \left[ (\overline{u}u+\overline{d}d) + (\overline{s}s) \right] \,.
\end{eqnarray}
Furthermore, using the Feynman-Hellman theorem, eq.~(\ref{fh_thm})
and with the hadron flavour expansion, eq.~(\ref{baryon_octet_linfit})
together with eq.~(\ref{M0_expan}) gives
\begin{eqnarray}
   \langle H| (\overline{u}u+\overline{d}d)^{\R} 
                - 2(\overline{s}s)^{\R} | H \rangle
    &=& {1 \over Z^{\NS}}\, c_H
\label{ren_octet}                                                      \\
   \langle H| (\overline{u}u+\overline{d}d)^{\R} 
                + (\overline{s}s)^{\R} | H \rangle
    &=& {1 \over Z^{\NS}}\, {M_0^{\prime} \over 1 + \alpha_Z} \,.
\label{ren_sing}
\end{eqnarray}
Eq.~(\ref{ren_octet}), the equation for the matrix element
of an octet operator, only involves $c_H$ (the hadron mass expansion
keeping the singlet quark mass constant), while eq.~(\ref{ren_sing}),
the matrix element of a singlet operator, only involves $M_0^\prime$
(occuring when changing the singlet quark mass). Eq.~(\ref{ren_octet})
also leads to eq.~(\ref{sigl_est}) as discussed in the introduction%
\footnote{The RHS of eq.~(\ref{ren_octet}) can be re-written as
$c_N/Z^{\NS} = 3A_1/Z^{\NS}$. Together with 
$M_\Xi + M_\Sigma -2M_N = - 9A_1\delta m_l = 3A_1(m_s^{\R} - m_l^{\R})/Z^{\NS}$ 
this gives  eq.~(\ref{sigl_est}). An alternative mass combination
that also picks out the $A_1$ coefficient is
$ M_\Xi - M_\Lambda = - 3A_1\delta m_l$.}.

Finally note that the quantities
\begin{eqnarray}
   (m_s - m_l) \langle H | (\overline{u}u + \overline{d}d)
                          -2 \overline{s}s | H \rangle \,, \quad
   (2 m_l + m_s) \langle H | (\overline{u}u + \overline{d}d)
                             + \overline{s}s | H \rangle \,,
\end{eqnarray}
are RGI, all $Z$ factors cancel when they are renormalised.
Linear combinations of these two quantities are also RGI
in particular the combination used previously of
$\sigma_l^{(H)} + \sigma_s^{(H)} = \sum_q m_q\langle H|\overline{q}q| H\rangle$.
However, $\sigma_l^{(H)}$ and $\sigma_s^{(H)}$ considered separately are
not RGI, see  eqs.~(\ref{renorm_mq}), (\ref{qqbar_ren}).
The renormalised quantities are mixtures of the two lattice quantities,
and $\alpha_Z$ is needed to relate lattice values to continuum values.
Refering back to Fig.~\ref{amN_plots} we see that the bare lattice
strange sigma term is much larger that the renormalised strange
sigma term, due to a cancellation between the two terms in
eq.~(\ref{qqbar_ren}).


\subsection{$\sigma$ equations}
\label{sigma_eqns}


Multiplying the renormalised quark mass, eq.~(\ref{renorm_mq}),
together with eqs.~(\ref{ren_octet}), (\ref{ren_sing})
(or more generally with eq.~(\ref{qqbar_ren})) we can find RGI
combinations (i.e.\ a form where the renormalisation constant
$Z^{\NS}$ cancels). In particular we find
\begin{eqnarray}
   \sigma_l^{(H)} - 2r \sigma_s^{(H)}
      &=& {3r \over 1 + 2r}(1+\alpha_Z)m_0c_H
\label{sigl_sigs_simul}                                      \\
   \sigma_l^{(H)} + r \sigma_s^{(H)}
      &=& {3r \over 1 + 2r}m_0 M_0^\prime(m_0) \,,
\label{sigs_sigl_simul}
\end{eqnarray}
where $r$ is the ratio of quark masses
\begin{eqnarray}
   r \equiv {m_l^{\R} \over m_s^{\R}} \,.
\end{eqnarray}
Thus we have to find the (fixed) coefficients $(1+\alpha_Z)m_0c_H$,
$m_0 M_0^\prime(m_0)$. We then determine the physical values
of the sigma terms by extrapolating to the point where the quark mass
ratio takes its physical value, i.e.\ $r = r^*$.

We observe that we have two simultaneous equations, which can be easily
solved to give%
\footnote{This leads to relations between the various sigma terms,
which we list in Appendix~\ref{approximate_rels} and where we also argue
that they are always approximately true.}
\begin{eqnarray}
   \sigma_l^{(H)} 
     &=& {r \over 1+2r}\,\left[ (1+\alpha_Z)m_0c_H 
                              + 2m_0M^{\prime}_0(m_0) \right]
                                                             \nonumber   \\
   \sigma_s^{(H)} 
     &=& {1 \over 1+2r}\,\left[ - (1+\alpha_Z)m_0c_H 
                              + m_0M^{\prime}_0(m_0) \right] \,.
\label{sigl_sigs}
\end{eqnarray}
We see that the smallness of $\sigma_l^{(H)}$ in comparison to
$\sigma_s^{(H)}$ is certainly guaranteed by the presence of
an additional $r$ in its numerator. As $\sigma_s^{(H)} > 0$ we
must also have $M^{\prime}_0(m_0) > (1+\alpha_Z)\max c_H$.
These coefficients are also sufficient to determine $y^{(H)\R}$,
as can be seen either directly from eq.~(\ref{sigl_sigs}) or
from eq.~(\ref{ren_sing}),
\begin{eqnarray}
   y^{(H)\R} = 2\, { -(1+\alpha_Z)m_0c_H + m_0M^{\prime}_0(m_0) \over
                   (1+\alpha_Z)m_0c_H + 2m_0M^{\prime}_0(m_0) } \,.
\label{yHR_formula}
\end{eqnarray}
Again, as seen in section~\ref{scalar_MEs}, $y^{(H)\R}$
only depends on gradients and not on the physical point.

It is now convenient to normalise the coefficients by $X_N$ 
so we now need to find the coefficients $(1+\alpha_Z)m_0c_H/X_N(m_0)$
and $m_0M^{\prime}_0(m_0)/X_N(m_0)$.


\subsection{Determination of the coefficients}
\label{det_coeff}


The hint for determining the coefficients from our lattice data
is given in section~\ref{scalar_MEs}, where we consider
gradients with respect to a renormalised or physical quantity
-- here taken as the pion mass. As in eq.~(\ref{baryon_octet_linfit})
we also have a similar expansion for the pseudoscalar octet,
\begin{eqnarray}
   M_\pi^2 = M_{0\,\pi}^2 + 2\alpha\delta m_l + O(\delta m_l^2) \,,
\label{pseudoscalar_octet_linfit}
\end{eqnarray}
(together with $M_K^2 = M_{0\,\pi}^2 - \alpha\delta m_l + O(\delta m_l^2)$,
$M_{\eta_s}^2 = M_{0\,\pi}^2 - 4\alpha\delta m_l + O(\delta m_l^2)$).
This gives a good representation of the data as can be seen
from Fig.~12 of \cite{bietenholz11a}.
Analogously to eq.~(\ref{XN_def}) we can define a flavour singlet quantity
\begin{eqnarray}
   X_\pi^2 = \third(2M_K^2+M_\pi^2) = M_{0\,\pi}^2 + O(\delta m_l^2) \,.
\end{eqnarray}
However, as well as eq.~(\ref{baryon_octet_linfit}), we have
the additional constraint from PCAC
\begin{eqnarray}
   M_\pi^2 = 2B_0^{\R}m_l^{\R} \,,
\label{mpi2_pcac}
\end{eqnarray}
(together with $M_K^2 = B_0^{\R}(m_l^{\R}+m_s^{\R})$,
$M_{\eta_s}^2 = 2B_0^{\R}m_s^{\R}$) which implies that
\begin{eqnarray}
   M_{0\,\pi}^2 = 2\alpha(1+\alpha_Z)\overline{m}\,, \qquad
   \alpha = B_0^{\R}Z^{\NS} \,.
\end{eqnarray}
If we now consider an expansion in the (physical) pion mass then
eliminating $\delta m_l$ between eq.~(\ref{baryon_octet_linfit})
and eq.~(\ref{pseudoscalar_octet_linfit}) gives
\begin{eqnarray}
   {M_H \over X_N} 
      = \left( 1 - \left[(1+\alpha_Z)m_0{c_H \over X_N}\right] \right)
        + \left[(1+\alpha_Z)m_0{c_H \over X_N}\right]\, {M_\pi^2 \over X_\pi^2} \,,
\label{mNoXN_mps2oXpi2}
\end{eqnarray}
from the point on the symmetric line $m_0 = \overline{m}$.
Thus if we plot $M_H / X_N$ versus $M_\pi^2 / X_\pi^2$ (holding the
singlet quark mass, $\overline{m}$ constant) then the gradient
immediately yields $(1+\alpha_Z)m_0c_H/X_N$. The only assumption is that
the `fan' plot splittings remain linear in $\delta m_l$ down to
the physical point. In 
Fig.~\ref{mpsO2omNOpmSigOpmXiOo32_mNOomNOpmSigOpmXiOo3_32x64_lin}
we show this plot giving the results
\begin{eqnarray}
   (1+\alpha_Z)m_0{3A_1 \over X_N} =
                   & 0.1899(55)\,,    & 0.2066(68)\,,
                                                            \nonumber \\
   (1+\alpha_Z)m_0{3A_2 \over X_N} =
                   & 0.03942(314)\,,  & 0.04164(431) \,,
\label{cHcoeff_num}
\end{eqnarray}
for $\kappa_0 = 0.12090$, $0.12092$ respectively.

\begin{figure}[htb]
   \begin{center}
      \includegraphics[width=8.00cm]
         {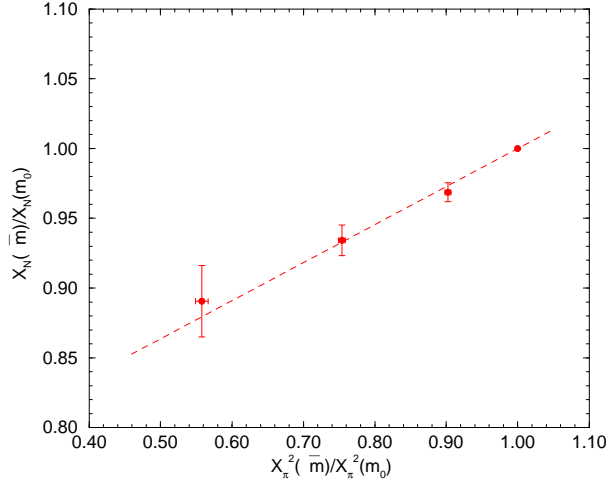}
   \end{center}
\caption{$X_N(\overline{m}) / X_N(m_0)$ versus
         $X_\pi^2(\overline{m}) / X_\pi^2(m_0)$ along the
         flavour symmetric line, together with the linear fit
         from eq.~(\protect\ref{XNoXN_Xpi2oXpi2}).}
\label{b5p50_XNoXNkp12090_Xpi2oXpi2kp12090}
\end{figure}

Alternatively on the flavour symmetric line, $m_l = \overline{m}$
(i.e.\ $\delta m_l = 0$), so varying $\overline{m}$ from a point $m_0$
gives
\begin{eqnarray}
   M_\pi^2(\overline{m})
      = M_{0\,\pi}^2(\overline{m})
      &=& M_{0\,\pi}^2(m_0) + M_{0\,\pi}^{2\,\prime}(m_0)(\overline{m} -m_0)
                                                            \nonumber \\
      &=& 2\alpha(1+\alpha_Z)[m_0 + (\overline{m} -m_0)] \,,
\label{MO2ps_expan}
\end{eqnarray} 
which gives $M_{0\,\pi}^{2\,\prime}(m_0) = 2\alpha(1+\alpha_Z)$.
So now eliminating $(\overline{m} - m_0)$ between
eqs.~(\ref{M0_expan}), (\ref{MO2ps_expan}) gives
\begin{eqnarray}
   {X_N(\overline{m}) \over X_N(m_0) }
     = \left( 1 - \left[{m_0 M_0^\prime(m_0) \over X_N(m_0)}\right] \right)
        + \left[{m_0 M_0^\prime(m_0) \over X_N(m_0)}\right] \,
           {X_\pi^2(\overline{m}) \over X_\pi^2(m_0)} \,.
\label{XNoXN_Xpi2oXpi2}
\end{eqnarray}
Again in a plot of $X_N(\overline{m}) / X_N(m_0)$ versus
$X_\pi^2(\overline{m}) / X_\pi^2(m_0)$ the gradient immediately
gives the required ratio $m_0M_0^\prime(m_0) / X_N(m_0)$. We have also
replaced $M_N$ by $X_N$ and $M_\pi^2$ by $X_\pi^2$ (which allows
us to use all the $32^3\times 64$ data available for a particular
$\overline{m}$). In Fig.~\ref{b5p50_XNoXNkp12090_Xpi2oXpi2kp12090}
we plot
$X_N(\overline{m}) / X_N(m_0)$ versus
$X_\pi^2(\overline{m}) / X_\pi^2(m_0)$. From eq.~(\ref{XNoXN_Xpi2oXpi2})
this gives
\begin{eqnarray}
   {m_0 M_0^\prime(m_0) \over X_N(m_0)} = 0.273(32) \,.
\label{mp_formula}
\end{eqnarray}

\begin{figure}[htb]
   \begin{center}
      \includegraphics[width=10.00cm]
         {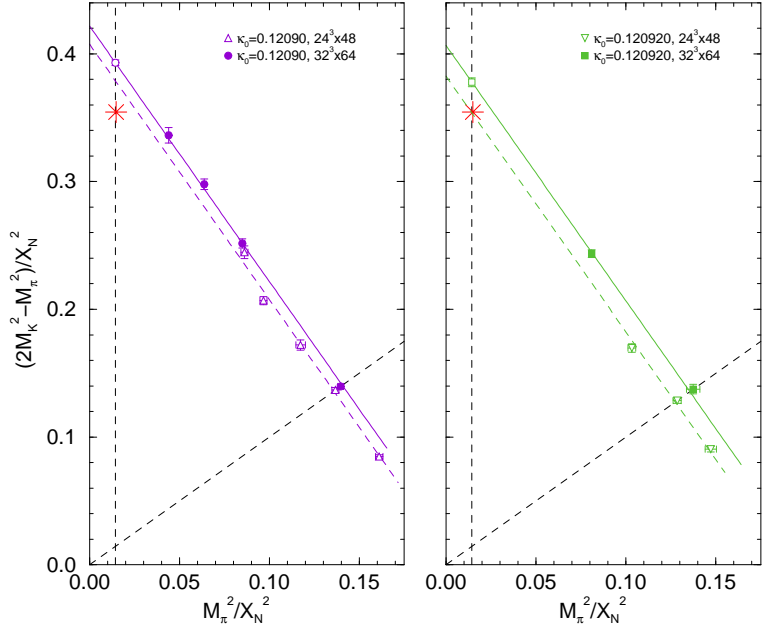}
   \end{center}
\caption{$(2M_K^2-M_\pi^2)/X_N^2$ versus $M_\pi^2 / X_N^2$
         for $\kappa_0 = 0.12090$ (left panel)
         and $\kappa_0 = 0.12092$ (right panel).
         The $32^3\times 64$ volume results are given by
         the filled symbols, while the $24^3\times 48$
         volume results are shown using empty triangles.
         The fit is given in eq.~(\protect\ref{const_mbar_fit}).
         Experimental points are denoted by (red) stars.}
\label{mps2oXn2_2mpsK2-mps2oXn2}
\end{figure}

Finally the quark mass ratio, $r$, must be estimated.
In Fig.~\ref{mps2oXn2_2mpsK2-mps2oXn2}
we plot $(2M_K^2-M_\pi^2)/X_N^2$ versus $M_\pi^2 / X_N^2$.
From eq.~(\ref{pseudoscalar_octet_linfit}) we have
\begin{eqnarray}
   {2 M_K^2 - M_\pi^2 \over X_N^2}
       = 3{M_{0\pi}^2 \over X_N^2} - 2{M_\pi^2 \over X_N^2} \,.
\label{const_mbar_fit}
\end{eqnarray}
As in section~\ref{flavour_sym_expan}, we see that for constant
$\overline{m}$ the data points lie on a straight line
(i.e.\ there is an absence of significant non-linearity).
Furthermore the gradient is fixed at $-2$. (Indeed
leaving the gradient as a fit parameter for the $\kappa_0 = 0.12090$
confirms that this gradient is very close to $-2$.)
Together with PCAC, eq.~(\ref{mpi2_pcac}) this gives 
the $x$-axis is proportional to $m_l^{\R}$ while
the $y$-axis is proportional to $m_s^{\R}$ and thus the
ratio gives $r$. Taking our physical scale to be defined
from $M_\pi^2/X_N^2|^*$ (i.e.\ from the $x$-axes of
Fig.~\ref{mps2oXn2_2mpsK2-mps2oXn2}) gives
\begin{eqnarray}
   {1 \over r^*} 
      = \left. {m_s^{\R} \over m_l^{\R}} \right|^*
          = \left\{ \begin{array}{cc}
                       27.28(16) & \kappa_0 = 0.12090 \\
                       26.23(24) & \kappa_0 = 0.12092 \\
                    \end{array}
            \right. \,.
\label{rstar_range}
\end{eqnarray}


\subsection{Curvature effects}
\label{curvature_effects}


What can we say about corrections to the linear terms?
The simple linear fit describes the data well, from the symmetric
point to our lightest pion mass, both along the
$\overline{m} = \mbox{const.}$ line and the flavour symmetric line.
To see qualitatively the possible influence of curvature we now compare
linear fits with quadratic fits. These will be used to estimate possible
systematic effects. We briefly discuss these effects here.

\begin{figure}[htb]
   \begin{minipage}{0.40\textwidth}
   \includegraphics[width=7.00cm]
      {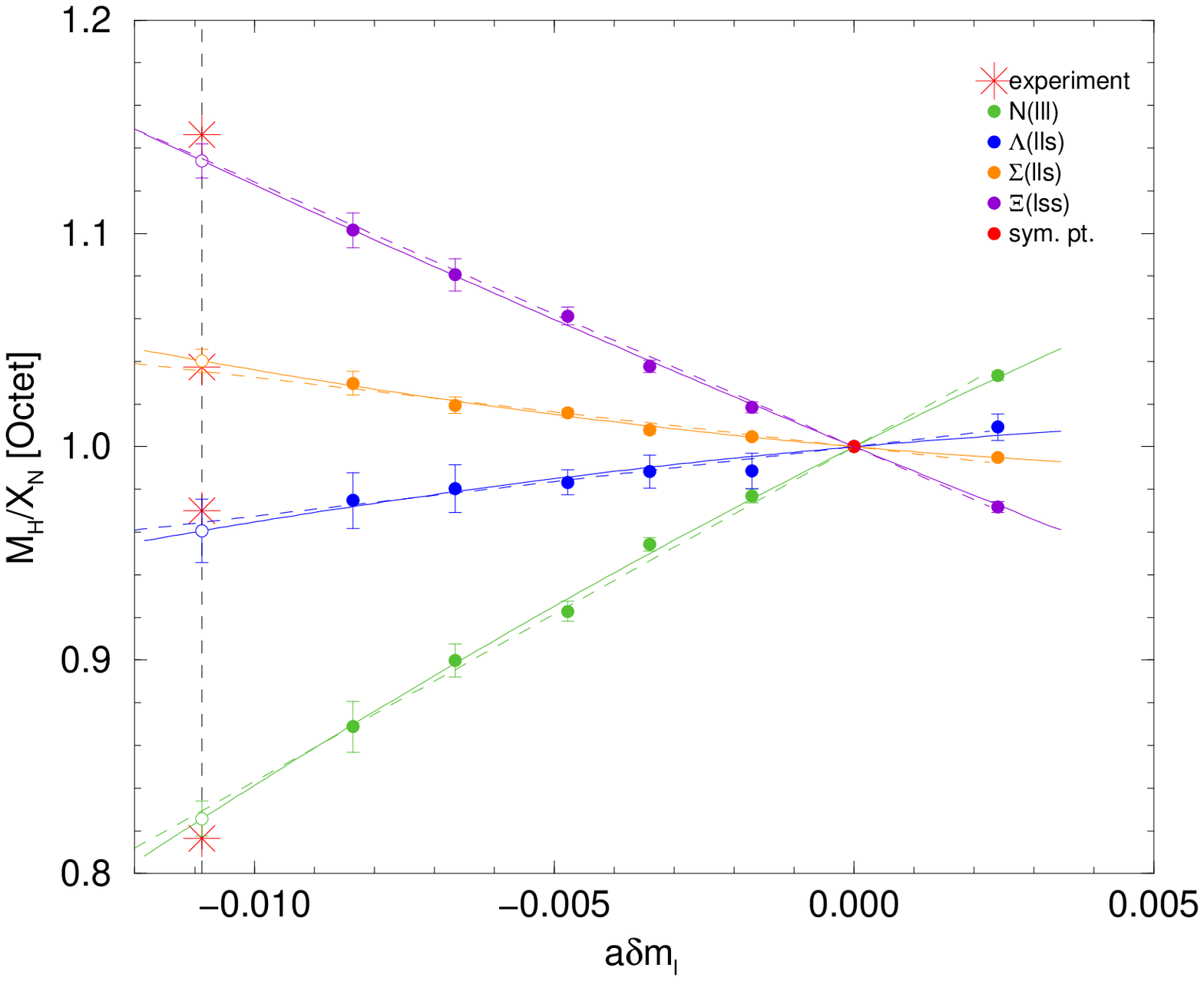}
   \end{minipage} \hspace*{0.10\textwidth}
   \begin{minipage}{0.40\textwidth}
   \includegraphics[width=7.00cm]
                  {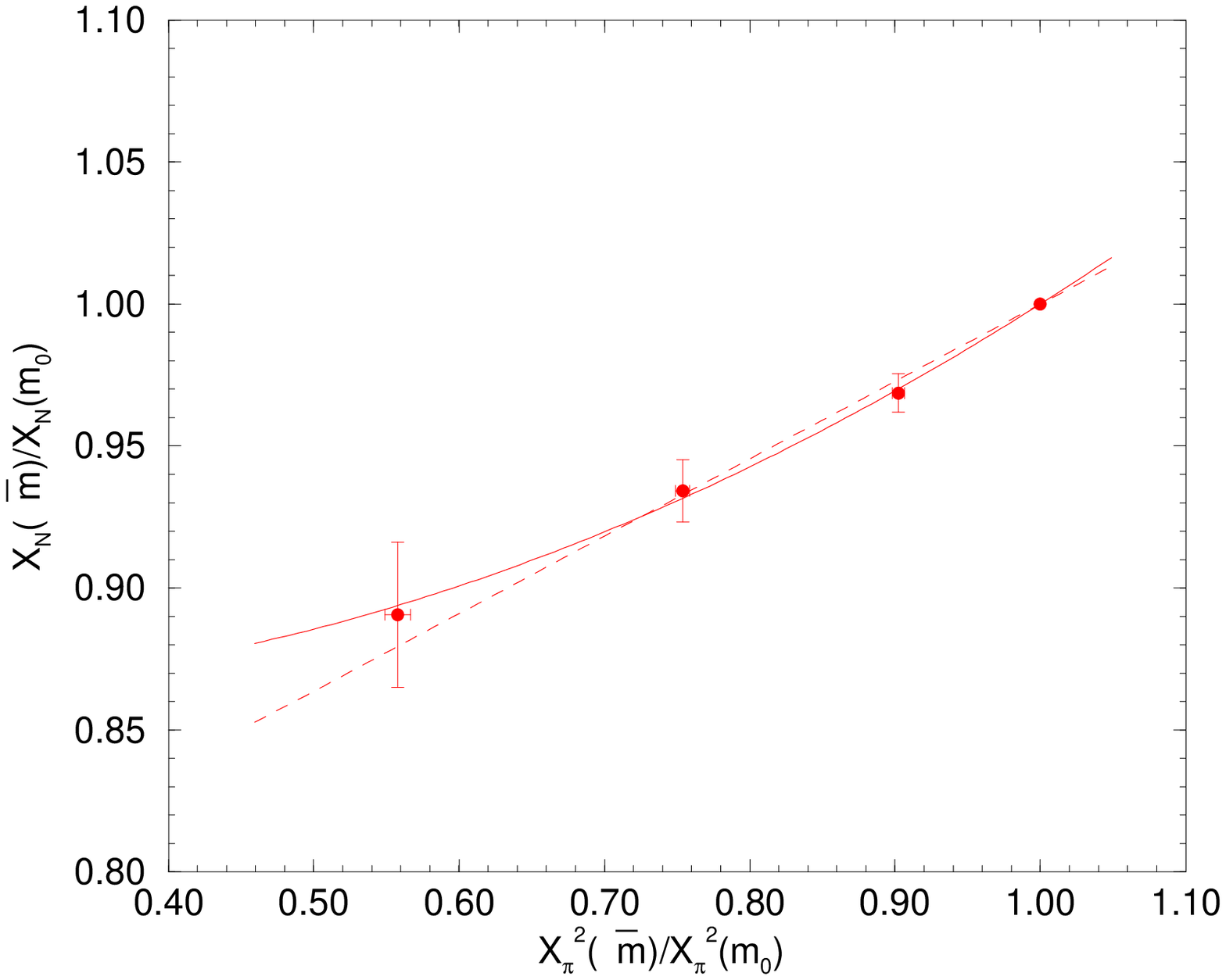}
   \end{minipage}
   \caption{Left panel: $M_H/X_N$ for $H = N$, $\Lambda$, $\Sigma$, $\Xi$
            against $a\delta m_l$ for initial point
            on the flavour symmetric line given by $\kappa_0 = 0.12090$
            together with the previous linear fit (dashed lines)
            and quadratic fit (solid lines). Other notation as in
            Fig.~\protect
            \ref{mpsO2omNOpmSigOpmXiOo32_mNOomNOpmSigOpmXiOo3_32x64_lin}.
            Right panel: $X_N(\overline{m}) / X_N(m_0)$ versus
            $X_\pi^2(\overline{m}) / X_\pi^2(m_0)$ along the
            flavour symmetric line, together with a linear fit
            from eq.~(\protect\ref{XNoXN_Xpi2oXpi2}) (dashed line) 
            and a quadratic fit (solid line).}
\label{b5p50_dml_mNOomNOpmSigOpmXiOo3_compar}
\end{figure}

In Fig.~\ref{b5p50_dml_mNOomNOpmSigOpmXiOo3_compar}
we compare the results of a quadratic fit and a linear fit, both for
the baryon mass fan plot and for $X_N(\overline{m}) / X_N(m_0)$.
In the left panel of the figure, we consider the  baryon mass fan
plot. The quadratic fit here uses all the data, \cite{bietenholz11a},
on both lattice sizes (in cases where results for two lattice sizes
are available, we used the larger lattice size only). 
The curvature terms here are small and statistically
compatible with zero.

The right panel of the figure shows a quadratic fit to the results
along the symmetric line. The curvature here is dominated by the
large error of the lightest point (which has a low statistic).
Thus we shall regard this fit as only giving an estimation of
the possible systematic error.

The results in the next section include systematic error estimates
from both these curvature sources combined in quadrature.
In Appendix~\ref{higher_order_effects} we give some more details.


\section{Results}
\label{results}


We can now numerically determine $y^{(H)\R}$ and $\sigma_l^{(H)}$,
$\sigma_s^{(H)}$.

We start with $y^{(H)\R}$. From eq.~(\ref{yHR_formula}), together
with eqs.~(\ref{cHcoeff_num}), (\ref{mp_formula}) and
eq.~(\ref{cHfoHinO}) gives the results in Table~\ref{table_y+sig}.
\begin{table}[htb]
   \begin{center}
      \begin{tabular}{lllll}
            &  \multicolumn{1}{c}{$N$}          &
               \multicolumn{1}{c}{$\Lambda$}    &
               \multicolumn{1}{c}{$\Sigma$}     &
               \multicolumn{1}{c}{$\Xi$}                               \\
         \hline
            \multicolumn{5}{c}{$\kappa_0 = 0.12090$}                   \\
         \hline
         $y^{(H)\R*}$      & 0.22(9)(15)     & 0.80(14)(28)  
                         & 1.23(20)(41)    & 2.14(38)(64)              \\
         $\sigma_l^{(H)*}\,\mbox{[MeV]}$ 
                          & 29(3)(4)       & 23(3)(4)    
                          & 20(3)(4)       & 16(3)(5)                  \\
         $\sigma_s^{(H)*}\,\mbox{[MeV]}$  
                          & 89(34)(59)     & 250(34)(68)  
                          & 334(34)(68)    & 453(34)(58)               \\
         \hline
            \multicolumn{5}{c}{$\kappa_0 = 0.12092$}                   \\
         \hline
         $y^{(H)\R*}$      & 0.18(9)(15)     & 0.79(14)(28)
                         & 1.25(20)(42)    & 2.30(42)(68)              \\
         $\sigma_l^{(H)*}\,\mbox{[MeV]}$ 
                          & 31(3)(4)       & 24(3)(4)
                          & 21(3)(4)       & 16(3)(4)                 \\
         $\sigma_s^{(H)*}\,\mbox{[MeV]}$ 
                          & 71(34)(59)     & 247(34)(69)
                          & 336(34)(69)    & 468(35)(59)               \\
         \hline 
      \end{tabular}
   \end{center}
\caption{Results for the baryon octet for $y^{(H)\R*}$,
         $\sigma_l^{(H)*}$, $\sigma_s^{(H)*}$ with $H = N$, $\Lambda$,
         $\Sigma$, $\Xi$ for $\kappa_0 = 0.12090$, $0.12092$.}
\label{table_y+sig}
\end{table}
The first error is the linear fit error (in this case dominated by
the error in eq.~(\ref{mp_formula})), while the second error
indicates possible effects from higher order terms,
as discussed in section~\ref{curvature_effects}.
We see that there is an order of magnitude increase in the
fraction of $\langle H| (\overline{s}s)^{\R} |H \rangle$ compared to
$\langle H| (\overline{u}u + \overline{d}d)^{\R} |H \rangle$
as we increase the strangeness content of the baryon from the
nucleon (no valence strange quarks) to the $\Xi$ (two valence strange
quarks).

Turning to the sigma terms themselves, from eq.~(\ref{sigl_sigs_simul})
we can find an indication of the magnitude of $\sigma^{(N)}_l$
as approximately (with $X_N = 1.1501\,\mbox{GeV}$),
\begin{eqnarray}
   \sigma_l^{(N)\,*} 
      \sim [22 \sim 25] + {\sigma_s^{(N)\,*} \over 13} \, \mbox{MeV}
      >  [22 \sim 25] \, \mbox{MeV} \,,
\label{sig_mbarconst_line}
\end{eqnarray}
(for $\kappa_0 = 0.12090$, $0.12092$ respectively). The last inequality
follows as obviously $\sigma_s^{(N)*} > 0$. Indeed this shows that a non-zero
$\sigma_s^{(N)*} > 0$ can only add a few $\mbox{MeV}$ to this result.

\begin{figure}[htbp]
   \begin{center}
      \includegraphics[width=5.00cm]{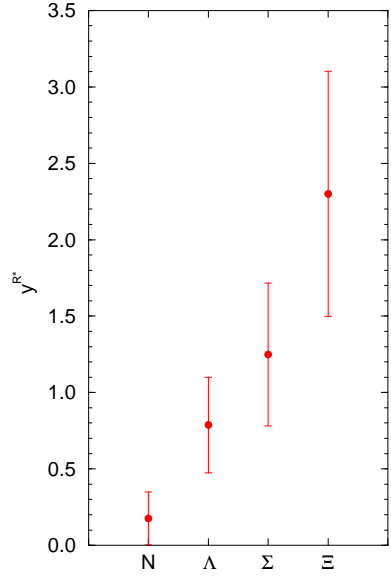}
   \end{center} 
   \caption{$y^{(H)\R*}$ for $H = N$, $\Lambda$, $\Sigma$, $\Xi$
            using the results from Table~\protect\ref{table_y+sig}
            for $\kappa_0 = 0.12092$.}
\label{yN}
\end{figure}

The results for $\sigma_l^{(H)*}$ and $\sigma_s^{(H)*}$ are also given
in Table~\ref{table_y+sig}. (Again the first error is the statistical
error, while the second systematic error is due to possible
quadratic effects.) While the data for $\kappa_0 = 0.12090$
is more complete than for $\kappa_0 = 0.12092$ (cf.\ the plots in
Fig.~\ref{mpsO2omNOpmSigOpmXiOo32_mNOomNOpmSigOpmXiOo3_32x64_lin})
and demonstrates linear behaviour, as the path starting at
$\kappa_0 = 0.12092$ is closer to the physical point
(cf.\ Fig.~\ref{mps2oXn2_2mpsK2-mps2oXn2}) we shall use
these values as our final values.
These results are illustrated in Fig.~\ref{yN} for $y^{(H)\R*}$
where $H = N$, $\Lambda$, $\Sigma$, $\Xi$.

\begin{figure}[htbp]
   \begin{center}
      \includegraphics[width=10.0cm]
            {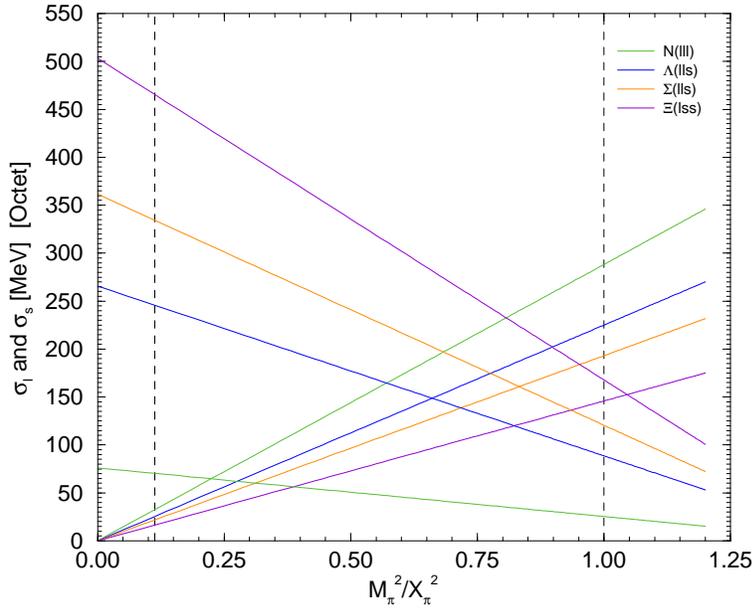}
   \end{center} 
   \caption{$\sigma_l^{(H)}$ (decreasing from the symmetric point
            $x=1$) and $\sigma_s^{(H)}$ (increasing) for $H = N$,
            $\Lambda$, $\Sigma$, $\Xi$ for $\kappa_0 = 0.12092$.
            The physical and symmetric lines are denoted by
            vertical dashed lines.}
\label{b5p50_mps2oXpi2_sigl+sigs}
\end{figure}

\begin{figure}[htbp]
   \begin{center}
      \includegraphics[width=10.0cm]
            {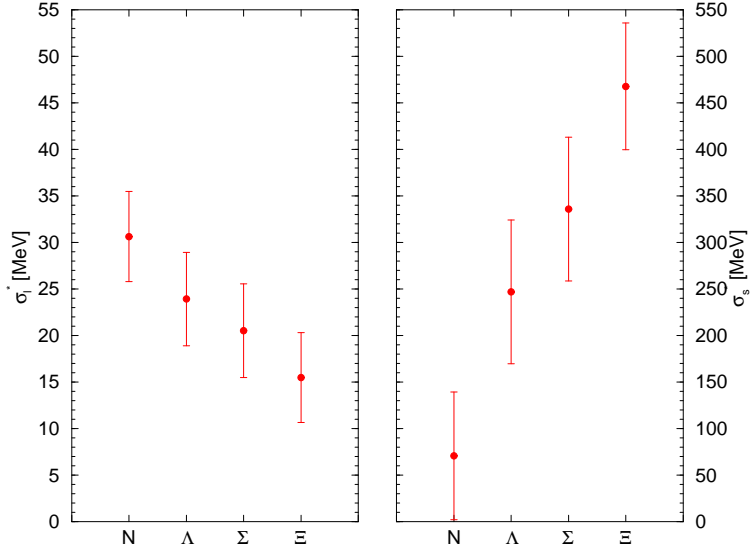}
   \end{center} 
   \caption{$\sigma_l^{(H)*}$ and $\sigma_s^{(H)*}$
            for $H = N$, $\Lambda$, $\Sigma$, $\Xi$ at the
            physical point for $\kappa_0 = 0.12092$.}
\label{b5p50_sigl+sigs_Xpi+XNscale}
\end{figure}

By varying $r$ in eq.~(\ref{sigl_sigs})%
\footnote{Using, for example, the results from the left panel
of Fig.~\ref{mps2oXn2_2mpsK2-mps2oXn2}, $r$ may be re-written as
\begin{eqnarray}
   r = { M_\pi^2/X_\pi^2 \over 3 - 2(M_\pi^2/X_\pi^2) } \,.
                                                         \nonumber
\end{eqnarray}}
, we plot in Fig.~\ref{b5p50_mps2oXpi2_sigl+sigs} 
$\sigma_l^{(H)}$ and $\sigma_s^{(H)}$ for the baryon octet,
$H = N$, $\Lambda$, $\Sigma$ and $\Xi$ from the symmetric point
(vertical dashed line at $x = 1$) to the physical point
(left vertical dashed line). $\sigma_l^{(H)}$ is rapidly
decreasing while $\sigma_s^{(H)}$ is increasing as we decrease
the quark mass. Also, as expected  $\sigma_l^{(H)}$ is largest 
for the nucleon, $N$, while $\sigma_s^{(N)}$ is the smallest.
Finally in Fig.~\ref{b5p50_sigl+sigs_Xpi+XNscale} we plot 
$\sigma_l^{(H)*}$, $\sigma_s^{(H)*}$ against $H = N$, $\Lambda$,
$\Sigma$ and $\Xi$, again using Table~\ref{table_y+sig}.


\section{Conclusions}


Keeping the average quark mass constant gives very linear `fan' plots
from the flavour symmetric point down to the physical point.
This implies that an expansion in the quark mass from the flavour
symmetric point will give information about the physical point.
In this article we have applied this to estimating the sigma terms
(both light and strange) of the nucleon octet. There has been no
use of a chiral perturbation expansion (indeed this is an opposite
expansion to the one used here, expanding about zero quark mass). 

Our results are given in section~\ref{results}
and we quote from there a value for the nucleon sigma terms of
\begin{eqnarray}
   \sigma_l^{(N)*} = 31(3)(4)\,\mbox{MeV} \,, \quad 
   \sigma_s^{(N)*} = 71(34)(59)\,\mbox{MeV} \,.
\end{eqnarray}
(The first error is the fit error while the second error indicates
possible effects from higher order terms in the flavour expansion.)
Note that expansions about the $SU(3)$ flavour line require
consistency between many QCD observables, here for example
not only for the baryon octet under consideration here,
but also for the pseudoscalar octet, and PCAC and the ratio of
the light to strange quark mass.

Of course there are several more avenues to investigate.
Numerically an increase in statistics for the masses
along the flavour symmetric line would reduce the
dominant error (both statistical and systematic)
and so directly help in decreasing the present errors.
Our approach here has been to emphasise linearity at
the expense (presently) of reaching exactly the physical point.
This can be addressed by interpolating between a small set
of constant $\overline{m}$ lines about the physical point.
Additionally the use of partial quenching will also help
to get closer to the physical pion mass. With more data,
a systematic investigation of quadratic quark mass terms
in the flavour expansion should be considered, to reduce the
systematic errors. Finally while the use of linear or quadratic
terms along the line of constant $\overline{m}$ is unproblematic,
so that it is unlikely that eq.~(\ref{sig_mbarconst_line})
will change by much, more subtle is the relation involving
$X(\overline{m})$ (i.e.\ the gradient when changing $\overline{m}$.)
For the example of clover fermions we have
$\tilde{g}^2(\overline{m}) = (1 + b_ga\overline{m})g^2$
which clearly does not change if $\overline{m} = \mbox{constant}$,
but will slightly change when $\overline{m}$ does. However this
is probably not a large effect (as $b_g$ seems small).
For a discussion of some aspects of this issue
see \cite{michale01a,babich10a}.


\section*{Acknowledgements}


The numerical configuration generation was performed using the
BQCD lattice QCD program, \cite{nakamura10a}, on the IBM
BlueGeneL at EPCC (Edinburgh, UK), the BlueGeneL and P at
NIC (J\"ulich, Germany), the SGI ICE 8200 at
HLRN (Berlin-Hannover, Germany) and the JSCC (Moscow, Russia).
We thank all institutions.
The BlueGene codes were optimised using Bagel, \cite{boyle09a}.
The Chroma software library, \cite{edwards04a},
was used in the data analysis. This work has been supported
in part by the EU grants 227431 (Hadron Physics2), 238353 (ITN STRONGnet)
and by the DFG under contract SFB/TR 55 (Hadron Physics from Lattice QCD).
JMZ is supported by STFC grant ST/F009658/1.



\appendix

\section*{Appendix}



\section{Some relations between the $\sigma$ terms}
\label{approximate_rels}


We discuss here some relations between the sigma terms
within a multiplet (here taken to be the baryon octet) 
which are exact within the linear case discussed here,
but which we might expect to always be approximately true.

The singlet relation eq.~(\ref{ren_sing}) or eq.~(\ref{sigs_sigl_simul})
is the same for every hadron. So in terms of sigma terms this
becomes 
\begin{eqnarray}
   \sigma_l^{(H)} + r \sigma_s^{(H)}
      \approx  \sigma_l^{(H^\prime)} + r \sigma_s^{(H^\prime)} \,.
\label{IDsing} 
\end{eqnarray}
At the flavour symmetric point it follows from group theory that
a singlet operator has the same value for every member of a multiplet,
so eq.~(\ref{IDsing}) must hold. But this can change if we move away
from the symmetric point. (We shall briefly discuss this at the end of
this section.)

We can find another collection of near identities by summing over 
a singlet combination of hadrons --- this can be either a singlet 
of $S_3$ or a singlet of $SU(3)$. If we do this, the expectation
values of $\overline{u}u$, $\overline{d}d$ and $\overline{s}s$ 
will be exactly equal at the flavour symmetry point, and stay again
nearly equal away from the symmetry point. By this argument we expect
\begin{eqnarray} 
   \sigma_l^{(\Lambda)} + \sigma_l^{(\Sigma)} 
      &\approx&  2 r \left( \sigma_s^{(\Lambda)} + \sigma_s^{(\Sigma)} \right)
                                                             \nonumber \\
   \sigma_l^{(N)} + \sigma_l^{(\Sigma)} +  \sigma_l^{(\Xi)} 
      &\approx&  2 r \left( \sigma_s^{(N)} + \sigma_s^{(\Sigma)} 
                                          +  \sigma_s^{(\Xi)}
                                             \right) \,.
\label{Xsig}
\end{eqnarray} 
(Again this relation, as with the other relations discussed here,
is exactly true for the linear case.)

Other relations come from the Gell-Mann--Okubo relation, 
\cite{gell-mann62a,okubo62a} in which the $27$-plet mass combination
is very small, 
\begin{eqnarray}
   2 M_N - 3 M_\Lambda - M_\Sigma + 2 M_\Xi \approx 0 \,,
\end{eqnarray}
for all values of $m_l, m_s$. In our approach,
its derivatives are also near zero. We therefore expect
\begin{eqnarray} 
   2 \sigma_l^{(N)} - 3\sigma_l^{(\Lambda)} - \sigma_l^{(\Sigma)} 
                                        + 2 \sigma_l^{(\Xi)}
      &\approx& 0
                                                             \nonumber \\
   2 \sigma_s^{(N)} - 3 \sigma_s^{(\Lambda)} - \sigma_s^{(\Sigma)} 
                                        + 2 \sigma_s^{(\Xi)}
      &\approx& 0 \,.
\label{sigl+sigs_rel}
\end{eqnarray} 
We obtain an even stronger version of these relations by taking the singlet
combination, proportional to
$(\overline{u}u + \overline{d}d)^{\R} + (\overline{s}s)^{\R}$,
\begin{eqnarray}
   2 \sigma_l^{(N)} - 3 \sigma_l^{(\Lambda)} - \sigma_l^{(\Sigma)} + 2\sigma_l^{(\Xi)}
      + r \left(  2 \sigma_s^{(N)} - 3 \sigma_s^{(\Lambda)} - \sigma_s^{(\Sigma)}
                  + 2\sigma_s^{(\Xi)} \right) 
   \approx 0 \,.
\label{stronger_sigs+sigl_rel}
\end{eqnarray}

There is also a relation between the sigma terms and the hadron masses,
\cite{ji94a} as the constants $A_1$ and $A_2$ which occur in the mass
splittings also occur in the leading order expressions for the sigma terms.
So there will be connections between masses and sigma terms. 
One particularly simple relation is 
\begin{eqnarray}
   M_H - \sigma_l^{(H)} - \sigma_s^{(H)} 
      \approx M_{H^\prime} - \sigma_l^{(H^\prime)} - \sigma_s^{(H^\prime)} \,.
\label{mass_relation}
\end{eqnarray}
(i.e.\ the baryon mass difference is closely accounted for by the sigma
terms.) For the linear case this is again exact, with this equation being
equal to $M_0(m_0) - m_0M_0^{\prime}(m_0)$ for all the octet baryons
(upon using eqs.~(\ref{baryon_octet_linfit}), (\ref{sigl_sigs})).
From eq.~(\ref{M0_expan}) we see that this is just the common
hadron mass in the chiral limit along the flavour symmetric line,
when $m_l = 0 = m_s$ or $\overline{m} = 0$.  $\sigma_l^{(H)}$ and
$\sigma_s^{(H)}$ can be thought of as that part of the hadron mass
which is due to $m_l$ and $m_s$ respectively. The remnant,
$ M_0(m_0) - m_0 M_0^\prime(m_0) $, is the part of the hadron mass
due to the quark and gluon kinetic energy, interaction
energy, etc., \cite{ji94a}, i.e.\ the part of the hadron mass which
is not due to the coupling with the Higgs vacuum expectation value.

We can use the higher order mass equations in \cite{bietenholz11a}
to estimate how well the relations in this section hold.
Most of the relations have violations proportional to the first
power of the $SU(3)$ breaking parameter, $\delta m_l$.
The corrections to eqs.~(\ref{IDsing}) and (\ref{Xsig})
and the first relation in eq.~(\ref{sigl+sigs_rel}) are
$O(m_l \delta m_l)$. The $\sigma_s$ relation in eq.~(\ref{sigl+sigs_rel})
has corrections $O(m_s \delta m_l)$. When we combine these two
relations to form eq.~(\ref{stronger_sigs+sigl_rel}),
the leading violation terms cancel, and we have a relation
with corrections $O(m_l \delta m_l^2)$.  The corrections to the
mass relation eq.~(\ref{mass_relation}) are $O(\overline{m}\delta m_l)$
and $O(\delta m_l^2)$.


\section{Higher order effects}
\label{higher_order_effects}


In this Appendix, we discuss a little more quantitatively the systematic
errors induced by the inclusion of the quadratic terms in the fit formulae.
We concentrate particularly on the nucleon sigma terms, $\sigma_l^{(N)}$
and $\sigma_s^{(N)}$.


\subsection{Curvature in the `fan' plot}
\label{curvature_fan_plot}


In Fig.~\ref{b5p50_dml_mNOomNOpmSigOpmXiOo3_compar}
we compare the results of a quadratic fit and a linear fit, both for
the baryon mass fan plot and on $\sigma_l^{(N)}$ and $\sigma_s^{(N)}$.
The quadratic fit uses all the data, \cite{bietenholz11a},
on both lattice sizes (in cases where results for two lattice sizes
are available, we used the larger lattice size only). Including curvature
terms in eq.~(\ref{baryon_octet_linfit}), \cite{bietenholz11a}, we have
$M_H = M_0 + c_H\delta m_l + b_H\delta m_l^2 + \ldots$.
Tracing through the analysis, we find the effect on eq.~(\ref{sigl_sigs})
is to replace 
\begin{eqnarray}
   c_H \to c_H + 2b_H\delta m_l \,.
\label{cH_change}
\end{eqnarray}
By comparing $c_H$
from the linear fit with $c_H + 2 b_H \delta m_l^* $ from the
quadratic fit, we can estimate the maximum possible change.

We use the data at $\kappa_0 = 0.12090$, because this is the case where
we have the most data, covering the largest range in quark mass splitting,
$\delta m_l$. In this case we have data covering about $3/4$ of the gap
from the symmetric point to the physical point, so we have the most
chance of seeing curvature effects if they are present.

\begin{figure}[htb]
   \begin{center}
      \includegraphics[width=5.650cm,height=6.75cm,angle=270]
         {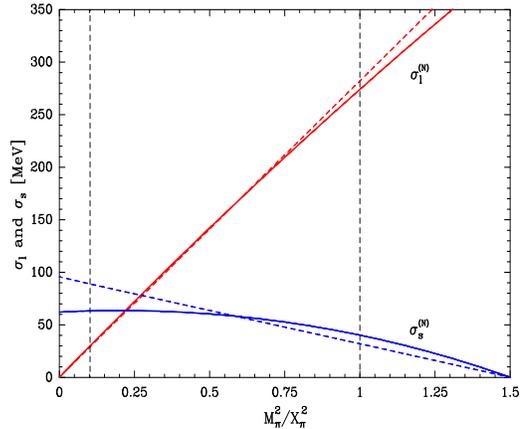}
   \end{center}
\caption{$\sigma_l^{(N)}$ (decreasing red lines from right to left)
         and  $\sigma_s^{(N)}$ (increasing blue lines from right to left)
         against $M_\pi^2/X_\pi^2$ using linear fits (dashed lines)
         and quadratic fits (solid lines) for $\kappa_0 = 0.12090$.}
\label{b5p50_quadsig3_kp12090}
\end{figure}

For the fan plot (left panel of
Fig.~\ref{b5p50_dml_mNOomNOpmSigOpmXiOo3_compar}), the curvature terms
are found to be small, and statistically compatible with zero curvature.
In Fig.~\ref{b5p50_quadsig3_kp12090} we compare the nucleon sigma terms
from the slopes of the two fits by using eq.~(\ref{sigl_sigs}) together
with eq.~(\ref{cH_change}). Again we see that the curvature effect is
very small in the case of $\sigma_l^{(N)}$, particularly at small $m_l$,
and much larger for $\sigma_s^{(N)}$. Can we explain this difference?

The slopes in the fan plot only effect the non-singlet matrix element,
the $c_H$ term in eq.~(\ref{sigl_sigs}). The curvature changes the
slope of the nucleon line by about $10\%$ at the physical point.
The non-singlet term in $\sigma_l^{(N)}$ is responsible for about $25\%$
of the quantity, so a $10\%$ change in slope translates to a $2.5\%$
change in $\sigma_l^{(N)}$. Putting in the actual slope change,
the final number we arrive at is a systematic uncertainty of about 
$1\,\mbox{MeV}$ in $\sigma_l^{(N)}$ coming from curvature in the fan plot.

The situation for $\sigma_s^{(N)}$ is different, the singlet and
non-singlet terms appear with opposite signs, so $\sigma_s^{(N)}$ is given
by the difference between two large quantities. Thus a $10\%$ change
in the non-singlet matrix element is leveraged into a $25\%$
change in $\sigma_s^{(N)}$. Repeating this procedure for the other
hadrons gives similar non-singlet uncertainties.


\subsection{Curvature along the symmetric line}
\label{curvature_symmetric_line}


We also use a linear fit to describe the baryon masses along the
symmetric line (the line with all three quark masses equal).
What is the effect of using a quadratic fit to determine the
slope along this line?

In the right panel of Fig.~\ref{b5p50_dml_mNOomNOpmSigOpmXiOo3_compar}
we compare a quadratic and linear fit to the symmetric baryon masses.
As before, the quadratic term is compatible with zero curvature.
Indeed the quadratic term is probably too large and is likely due
to having a short lever arm and low statistics at the lightest point
rather than to be a real effect. (Also we would expect that chiral
perturbation theory would predict a downward curve.)

Feeding these values into eq.~(\ref{sigl_sigs}) gives an estimate
of the possible effect of quadratic terms, due to curvature along
the symmetric line, which we will include in our final error estimate.
This curvature effect is the same for every hadron, giving an uncertainty
$\sim 4\,\mbox{MeV}$ for $\sigma_l$ and $\sim 55\,\mbox{MeV}$ for
$\sigma_s$. However because the shift is universal, this does not effect
splittings, so the systematic error in $\sigma_l^{(H)} - \sigma_l^{(H^\prime)}$
is still given by the $\sim 1\,\mbox{MeV}$ value of the previous subsection.
For $y^{(H)\R}$, using the first equation in eq.~(\ref{sigs_est}) gives
percentage changes in $y^{(N)\R}$ of $60\%$ and $30\%$ for
$y^{(\Lambda)\R}$, $y^{(\Sigma)\R}$ and $y^{(\Xi)\R}$.


\section{Hadron Masses}
\label{hadron_masses}


We collect here in 
Tables~(\ref{table_run_M_O_ksym}) -- (\ref{table_run_M_BOoX})
numerical values for the meson pseudoscalar octet
and baryon octet, not given in \cite{bietenholz11a}. (All the data sets
used here are over $\sim 2000$ configurations for the $24^3\times 48$ volumes
and $\sim 1500-2000$ configurations for the $32^3\times 64$
volumes except for $\kappa_0=0.12099$ which has $\sim 500$ configurations.)
Errors are from a bootstrap analysis.


\begin{table}[htbp]
   \begin{center}
      \begin{tabular}{lll}
\multicolumn{1}{c}{$\kappa_0$}                               &
\multicolumn{1}{c}{$aM_\pi$}                                  &
\multicolumn{1}{c}{$aM_N$}                                   \\
         \hline
\multicolumn{3}{c}{$32^3\times 64$}                          \\
         \hline
         0.120920 & 0.1647(4)  & 0.4443(59)                  \\
         \hline
      \end{tabular}
   \end{center}
\caption{Additional result for the pseudoscalar octet mesons and
         octet baryons along the flavour symmetric line: $aM_\pi$, $aM_N$,
         for $(\beta, c_{sw}, \alpha) = (5.50, 2.65, 0.1)$.}
\label{table_run_M_O_ksym}
\end{table}


\begin{table}[htbp]
   \begin{center}
      \begin{tabular}{llll}
\multicolumn{1}{c}{$(\kappa_l,\kappa_s)$}                     &
\multicolumn{1}{c}{$aM_\pi$}                                  &
\multicolumn{1}{c}{$aM_K$}                                    &
\multicolumn{1}{c}{$aM_{\eta_s}$}                             \\
         \hline
\multicolumn{4}{c}{$24^3\times 48$}                           \\
         \hline
  (0.120870, 0.121020) & 0.1804(8)  & 0.1621(10) & 0.1407(12) \\
  (0.120980, 0.120800) & 0.1545(9)  & 0.1775(8)  & 0.1976(7)  \\
         \hline
      \end{tabular}
   \end{center}
\caption{Additional results for the pseudoscalar octet mesons:
         $aM_\pi$, $aM_K$ and $aM_{\eta_s}$ for
         $(\beta, c_{sw}, \alpha) = (5.50, 2.65, 0.1)$
         where $\kappa_0 = 0.12092$.}
\label{table_run_M_MO_ksym12092}
\end{table}


\begin{table}[htbp]
   \begin{center}
      \begin{tabular}{lllll}
\multicolumn{1}{c}{$(\kappa_l,\kappa_s)$}                     &
\multicolumn{1}{c}{$aM_N$}                                    &
\multicolumn{1}{c}{$aM_\Lambda$}                               &
\multicolumn{1}{c}{$aM_\Sigma$}                                &
\multicolumn{1}{c}{$aM_\Xi$}                                  \\
         \hline
\multicolumn{5}{c}{$24^3\times 48$}                           \\
         \hline
  (0.120870, 0.121020) & 0.4812(40) & 0.4721(62) & 0.4672(48) & 0.4618(58) \\
  (0.120980, 0.120800) & 0.4668(61) & 0.4773(62) & 0.4838(47) & 0.4909(41) \\
         \hline
      \end{tabular}
   \end{center}
\caption{Additional results for the octet baryons: $aM_N$, $aM_\Lambda$,
         $aM_\Sigma$ and $aM_\Xi$ for
         $(\beta, c_{sw}, \alpha) = (5.50, 2.65, 0.1)$
         where $\kappa_0 = 0.12092$.}
\label{table_run_M_BO_ksym12092}
\end{table}


\begin{table}[htbp]
   \begin{center}
      \begin{tabular}{lllll}
\multicolumn{1}{c}{$(\kappa_l,\kappa_s)$}                     &
\multicolumn{1}{c}{$M_N/X_N$}                                 &
\multicolumn{1}{c}{$M_\Lambda/X_N$}                           &
\multicolumn{1}{c}{$M_\Sigma/X_N$}                            &
\multicolumn{1}{c}{$M_\Xi/X_N$}                               \\
         \hline
\multicolumn{5}{c}{$24^3\times 48$}                           \\
         \hline
  (0.120870, 0.121020) & 1.024(3)   & 1.004(9)   & 0.9939(17) & 0.9824(34) \\
  (0.120980, 0.120800) & 0.9715(33) & 0.9934(95) & 1.007(2)   & 1.022(3)   \\
         \hline
\multicolumn{5}{c}{$32^3\times 64$}                           \\
         \hline
  (0.121050, 0.120661) & 0.9167(40) & 0.9872(46) & 1.017(2)   & 1.066(3)   \\
         \hline
      \end{tabular}
   \end{center}
\caption{Additional ratio results for the octet baryons:
         $M_N/X_N$, $M_\Lambda/X_N$, $M_\Sigma/X_N$ and $M_\Xi/X_N$ for
         $(\beta, c_{sw}, \alpha) = (5.50, 2.65, 0.1)$
         where $\kappa_0 = 0.12092$.}
\label{table_run_M_BOoX}
\end{table}




\end{document}